%% file: main.tex
\documentclass[twocolumn]{aastex63}

\usepackage[T1]{fontenc}
\shorttitle{RR Lyrae stars in the Pal 5 stream}
\shortauthors{Price-Whelan et al.}

\newcommand{\changes}[1]{#1}

\newcommand{\clderr}{\ensuremath{20.6 \pm 0.2~\kpc}}
\newcommand{\NRRL}{27}     

\newcommand{\NRRab}{15}    
\newcommand{\NRRc}{12}     
\newcommand{\NRRcl}{10}     
\newcommand{\NRRtails}{17} 

\input{defs.tex}

\begin{document}\sloppy\sloppypar\raggedbottom\frenchspacing 

\title{Kinematics of the Palomar 5 stellar stream from RR Lyrae stars}

\author[0000-0003-0872-7098]{Adrian~M.~Price-Whelan}
\affiliation{Center for Computational Astrophysics, Flatiron Institute,
             Simons Foundation, 162 Fifth Avenue, New York, NY 10010, USA}
\affiliation{Department of Astrophysical Sciences,
             Princeton University, Princeton, NJ 08544, USA}
\email{aprice-whelan@flatironinstitute.org}
\correspondingauthor{Adrian M. Price-Whelan}

\author[0000-0002-6330-2394]{Cecilia~Mateu}
\affiliation{Departamento de Astronom\'ia, Facultad de Ciencias, Universidad de la Rep\'ublica, Igu\'a 4225, 14000, Montevideo, Uruguay}

\author[0000-0003-0293-503X]{Giuliano~Iorio}
\affiliation{Institute of Astronomy, University of Cambridge, Madingley Road, Cambridge CB3 0HA, UK}

\author[0000-0003-0256-5446]{Sarah~Pearson}
\affiliation{Center for Computational Astrophysics, Flatiron Institute,
             Simons Foundation, 162 Fifth Avenue, New York, NY 10010, USA}

\author[0000-0002-7846-9787]{Ana Bonaca}
\affiliation{Center for Astrophysics | Harvard \& Smithsonian, 60 Garden St, Cambridge, MA 02138, USA}

\author[0000-0002-0038-9584]{Vasily~Belokurov}
\affiliation{Institute of Astronomy, University of Cambridge, Madingley Road, Cambridge CB3 0HA, UK}

\begin{abstract}
Thin stellar streams, formed from the tidal disruption of globular clusters, are important gravitational tools, sensitive to both global and small-scale properties of dark matter.
The Palomar 5 stellar stream (Pal 5) is an exemplar stream within the Milky Way: Its $\sim 20\degr$ tidal tails connect back to the progenitor cluster, and the stream has been used to study the shape, total mass, and substructure fraction of the dark matter distribution of the Galaxy.
However, most details of the phase-space distribution of the stream are not fully explained, and dynamical models that use the stream for other inferences are therefore incomplete.
Here we aim to measure distance and kinematic properties along the Pal 5 stream in order to motivate improved models of the system. 
We use a large catalog of RR Lyrae-type stars (RRLs) with astrometric data from the \Gaia\ mission to probabilistically identify RRLs in the Pal 5 stream.
RRLs are useful because they are intrinsically-luminous standard candles and their distances can be inferred with small relative precision ($\sim3\%$).
By building a probabilistic model of the Pal 5 cluster and stream in proper motion and distance, we find \NRRL\ RRLs consistent with being members of the cluster (\NRRcl) and stream (\NRRtails).
Using these RRLs, we detect gradients in distance and proper motion along the stream, and provide an updated measurement of the distance to the Pal 5 cluster using the RRLs, $d = \clderr$.
We provide a catalog of Pal 5 RRLs with inferred membership probabilities for future modeling work.
\end{abstract}

\keywords{globular clusters: individual: Palomar 5 ---
stars: variables: RR Lyrae ---
Galaxy: halo ---
Galaxy: structure}

\section{Introduction} \label{sec:intro}

Globular clusters are destroyed as they orbit within the Milky Way.
The $\sim$150 bound globular clusters we presently see throughout the Galaxy \citep{Harris:2010} are therefore thought to be surviving relics of a much larger initial population, most of which were destroyed (primarily) through a combination of relaxation / evaporation and gravitational shocking from the disk and bulge \citep[e.g.,][]{Gnedin:1997, Gnedin:2014}.
As stellar systems are destroyed---i.e., as stars are tidally stripped from their progenitor---the tidal debris forms tails of matter that both lead and trail the remnant cluster, producing a stellar stream that may persist even after the progenitor is fully destroyed \citep[e.g.,][]{Johnston:1996}.

Stellar streams are useful objects because they almost \citep[see][]{Sanders:2013} delineate the past and future trajectory of their progenitor system, offering information about orbits that can be used to infer the mass distribution from the single kinematic snapshot of the Galaxy that we are afforded \citep[e.g.,][]{Johnston:1999, Sanders:2013, PriceWhelan:2014, Bonaca:2018, Malhan:2019, Erkal:2019}. 
Many globular cluster stellar streams have been discovered over the last few decades \citep[see, e.g.,][]{Grillmair:2016, Shipp:2018}, in large part because of large-area, deep, multi-band imaging surveys such as the Sloan Digital Sky Survey \citep[SDSS;][]{York:2000}, Pan-STARRS PS1 survey \citep{Chambers:2016}, and Dark Energy Survey \citep{DES:2016}.
A subset of the known streams have been precisely characterized and further studied \citep[e.g.,][]{PriceWhelan:2018, Malhan:2018, Shipp:2019} using kinematic data from the recent data release 2 (\DR{2}) of the \Gaia\ mission \citep{Gaia:2016, Gaia:2018}.
Of the currently known $\sim$60 candidate stellar streams found throughout the Milky Way, few have known progenitor systems.
Conversely, of the $\sim$150 known globular clusters, only $\lesssim 20\%$ have purported tidal tails \citep[e.g.,][]{Leon:2000, Kundu:2019}, but even these are mostly low density and tenuous \citep[as might be expected, e.g.,][]{Balbinot:2018}.
A prominent exception to these statements is the Palomar 5 (Pal 5) globular cluster and stream.

The Pal 5 stream---at a heliocentric distance of $d \sim 20~\kpc$---was discovered using early SDSS imaging, which reached well below the main sequence turnoff of the stream over a large area surrounding the cluster \citep{Odenkirchen:2001, Rockosi:2002}.
Since then, radial velocities have been measured for a handful of giant branch stars in the cluster and along the stream \citep{Odenkirchen:2002, Odenkirchen:2009, Ibata:2017}, and deeper imaging data has been used to map the tails at higher signal-to-noise \citep{Bernard:2016, Ibata:2016, Bonaca:2019}.
The detailed but still limited kinematic data for Pal 5 has made it a canonical example of tidal destruction and a useful tool for constraining Galactic structure.
For example, using the sky track and radial velocity information, the stream has been used to measure the enclosed mass of the Milky Way \citep{Kuepper:2015, Bovy:2016, Dai:2018}.
Using deep photometry, the density variations along the stream have been used to place limits on the abundance of massive substructures in the Galactic halo \citep{Erkal:2017}.

However, a number of peculiar aspects of the stream still remain unexplained in detail.
For example, the leading and trailing tails have significantly different total star counts and lengths \citep[both different by a factor of $\sim$2 between leading and trailing;][]{Dehnen:2004, Bernard:2016}, and the stream track, width, and density vary significantly over the full extent of the stream \citep{Ibata:2016, Bonaca:2019}.
Given that Pal 5's pericenter is well within the Galactic disk ($r_{\textrm{peri}} \sim 7$--$8~\kpc$),
these features could be a sign of perturbations from molecular clouds \citep[e.g.,][]{Amorisco:2016}, interaction with the Galactic bar \citep[e.g.,][]{Pearson:2017}, signatures of past dark matter subhalo encounters \citep[e.g.,][]{Erkal:2017}, or some combination of all of these effects.
Mapping the stream in all six phase-space dimensions---i.e., combining deep photometry with radial velocity, distance, and proper motion measurements---will enable and motivate improved dynamical models of the stream and more precise constraints on the Galactic mass distribution \citep[e.g.,][]{PriceWhelan:2013}.
However, with \Gaia\ \DR{2}, at its present distance, main sequence stars are barely detected and have large astrometric uncertainties.
Other stellar tracers (e.g., giant branch stars) are not numerous enough or precise enough distance indicators to provide useful relative distance information along the stream.

RR Lyrae stars (RRLs) are pulsating horizontal branch giants intrinsically more luminous than main sequence turn-off stars. Although about $\sim 10,000$ time less numerous than main sequence stars, RRLs are extremely useful probes of the Galactic halo as they are well-established as standard candles; they can be observed to large distances \citep{Medina:2018,Sesar:2017c}; and are reliably identified based on their photometric variability, with little to no contamination from other types of variable stars \citep[e.g.,][]{Holl:2018,Drake:2017,Mateu:2012}. Their distances can be estimated based on photometric data alone, with errors $\sim8\%$ in the optical, and can be as low as 3\% in the infrared \citep{Neeley:2017}. These properties have made RRLs a key tracer of structure and substructure across the Galaxy, having been used to trace the density profile of the halo, thick disk and bulge \citep[e.g.,][and reference therein]{Iorio:2018,Mateu:2018b,Kunder:2008}, to identify new streams and accretion events \citep[e.g.]{Duffau:2006,Sesar:2010,Mateu:2018,Iorio:2019} and even to argue against the extragalactic origin of overdensities in the disk \citep{Mateu:2009, Price-Whelan:2015}.

Previous studies have found RRLs in Pal5's tails: the first two were found by \citet{Vivas:2001} in the QUEST survey, and more recently \citep{Ibata:2017} found about ten new RRLs using a catalog of PS1 candidate RRLs from \citet{Hernitschek:2016}. At its current distance, Pal~5's horizontal branch is bright enough ($\G\sim17.3$) to be observed by \Gaia. Therefore, combining \Gaia~DR2 proper motions with the newer PS1 RRL catalog from \citet{Sesar:2017b}---with improved classification probabilities and light curve periods---enables us to make the first precise distance and proper motion measurements along the full extent of the Pal~5 stream. Both will prove to be key ingredients for dynamical modelling of the Pal~5 stream.

\begin{figure*}[t!]
\begin{center}
\includegraphics[width=0.95\textwidth, trim=0 20px 0 0]{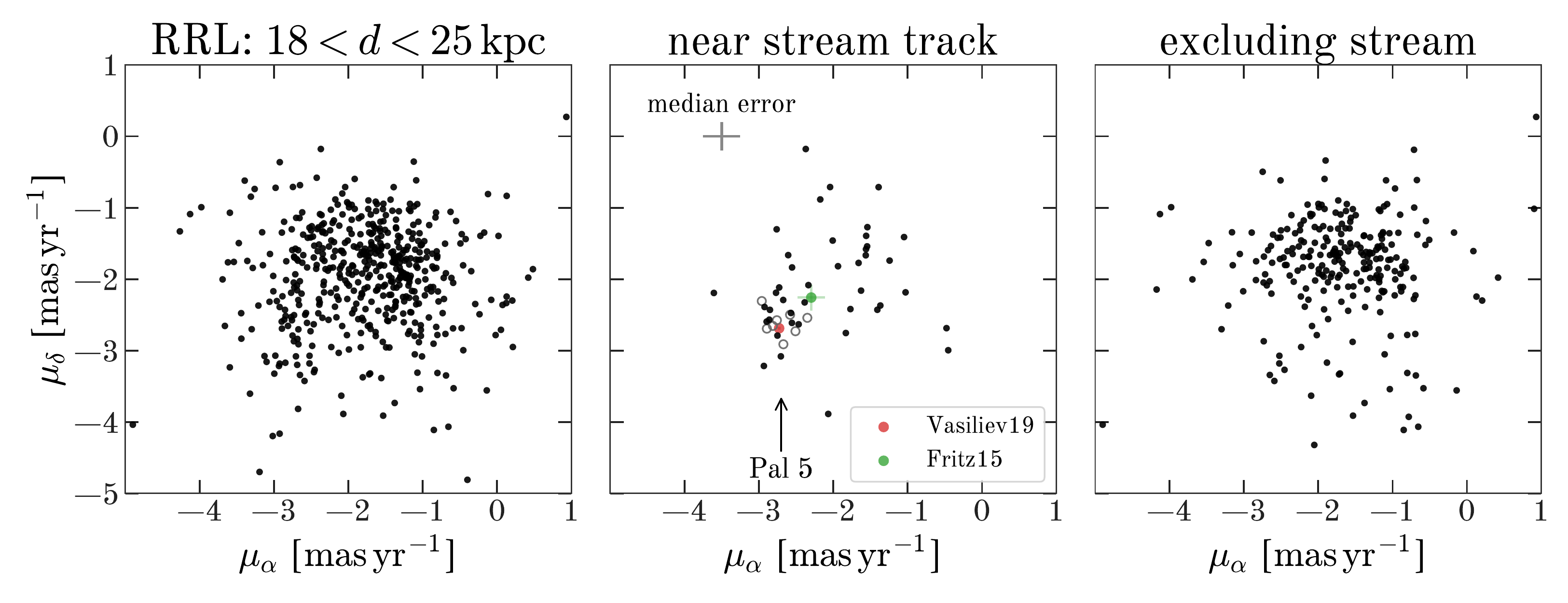}
\caption{\textbf{Left}: Proper motions of RRLs in the region $215^\circ < \alpha < 255^\circ$, $-15^\circ < \delta < 10^\circ$ with distances in the range $18 < d < 25~\kpc$.
The distance selection here is purely illustrative: We do not use a distance cut in our membership model.
\textbf{Middle}: The same, but for RRLs within $1\degr$ of the mean sky track of Pal 5 \citep{Bonaca:2019}.
\changes{The open circle markers indicate RRLs that lie within the Jacobi radius of the cluster (see \sectionname~\ref{sec:results}) that pass this simple selection.}
The over-density of stars near $(-2.7, -2.7)$ is the cluster and stream, and the two colored points show the Pal 5 cluster proper motions from \citet{Vasiliev:2019} and \citet{Fritz:2015}, respectively.
\textbf{Right}: The same, but for RRLs excluding $2\degr$ around the Pal 5 sky track.}
\label{fig:pm}
\end{center}
\end{figure*}

In this Article, we construct a probabilistic model for kinematic properties of RRLs in the vicinity of the Pal 5 stream to determine membership probabilities and thus map the distance and proper motion trends in the stream.
In \sectionname~\ref{sec:data}, we describe the source RRL catalog we use to search for Pal 5 members.
In \sectionname~\ref{sec:membership}, we implement a probabilistic membership model to identify candidate members of the stream using distance and astrometric information.
In \sectionname~\ref{sec:results}, we discuss the population and kinematics of the RRL members found and use these to provide an estimate of the system's stellar mass and luminosity.
We present our conclusions in \sectionname~\ref{sec:conclusions}.

\section{Data} \label{sec:data}

The PanSTARRS-1 (PS1) catalog of RRLs \citep{Sesar:2017b} contains 229K stars spanning $\sim$3/4 of the sky (all sky north of declination $\delta > -30\degr$), with multi-epoch $grizy$ data up to a limiting magnitude $r\sim21$ (corresponding to $G \sim 21$ for the typical colors of RRLs). These stars were identified as RRLs using machine-learning methods based on a template fitting algorithm that can find pulsation periods overcoming the sparse sampling of the PS1 survey ($\sim12$ epochs per filter).

Two all-sky RRL catalogs have also been released as part of \Gaia~\DR{2} \citep[VariClassifier and Specific Object Studies][]{Holl2018, Rimoldini2018, Clementini2018}. At present, however, these catalogs are subject to more significant and spatially-varying incompleteness in this particular area of the sky due to the \Gaia\ scanning law \citep[see][though many of these issues will improve with future \Gaia\ data releases]{Rimoldini2018}.

Out of the 229K RRLs in the PS1 catalog, $\sim60$K are also reported as RRLs by \Gaia~\DR{2}, the vast majority of which (49K) are part of the subset defined by \citet{Sesar:2017b} as \emph{bona fide}. This subset is comprised of 61K RRLs above classification score thresholds set by \citet{Sesar:2017b} to ensure its high purity ($>90$\%) and completeness ($>80$\%). An extra $\sim6$K stars with lower classification scores in PS1 were also found to be RRLs by \Gaia~\DR{2} and have reported pulsation periods in the SOS catalog. Most of these RRLs have matching classifications and consistent periods in the two surveys, which increases our confidence that these are true RRLs.

For our present work, therefore, we use the superset of 68\,254 stars consisting of the 61\,795 \emph{bona fide} PS1 RRLs, plus the 6\,459 non-\emph{bona fide} PS1 stars also reported in the SOS \Gaia~\DR{2} catalog. Out of these, 68\,085 RRLs (99.8\%) have astrometric information available in \Gaia~\DR{2} and all---by construction---have photometric data and pulsation periods from PS1.

\citet{Sesar:2017b} provide distances to the RRLs based on an $i$-band Period-Luminosity-Metallicity (PLZ) relation.
This has two key advantages when compared to optical bands: The PLZ relation in the $i$-band only weakly depends on metallicity, and the impact of dust extinction is less severe at longer wavelengths.
However, distances reported for the \rrc~stars in Table~5 of \citet{Sesar:2017b} have a systematic offset with respect to the \typeab~stars.
This offset comes from having applied the same PLZ to \typeab~and \typec~stars.
This is corrected by using the `fundamentalized' period in the PLZ relation for the \rrc~stars, which is computed as $\log{P_F} = \log P + 0.126$ \citep[following][]{Braga2016}.
In what follows, we use the $i$-band PLZ distances for the RRLs.
\citet{Sesar:2017b} estimate their overall distance precision to be 3\%, of which $\sim2$\% and $\sim1$\% correspond to systematic and random uncertainties, respectively.

We cross-match the full PS1 RRL catalog with \Gaia~DR2 \citep{Gaia:2018}, using a 1" sky separation tolerance, in order to retrieve astrometric information for these stars.
Although not all PS1 RRLs have been identified as RRLs from \Gaia\ photometry, all of the PS1 RRLs are present in the main point source catalog.
\figurename~\ref{fig:pm} demonstrates that Pal 5 \changes{cluster and stream} members appear in the RRL catalog: This figure shows the \Gaia\ \DR{2} proper motions for all RRLs in the selected sky window (indicated in the caption) within the distance range $18 < d < 25~\kpc$, then the same for RRLs in a tight selection around the observed stream track (see \sectionname~\ref{sec:membership} for details), and then finally the same for RRLs \emph{excluding} the stream region.

\citet{Sesar:2017b} provide an estimate of their catalog's completeness as a function of distance, at high galactic latitude, based on simulations. They estimate the mean completeness to be 92\% for \rrab~and 79\% for \rrc, up to a distance of 40~kpc, well beyond the distance of Pal~5 and its tails. At the average ratio of 3 \rrc~stars per every 10 \rrab~\citep{Layden:1995}, this represents a mean completeness of 89\%. We can also produce an estimate per line of sight using the methodology described in \citet{Rybizki:2018}, which requires two independent catalogs to assess the completeness of both in a probabilistic manner. Comparing PS1 against the \Gaia~DR2 catalogue, comprised of the outer join of the VariClassifier \citep{Holl2018,Rimoldini2018} and Specific Objects Study \citep{Clementini2018} catalogs, we find a median completeness of 92\% with a standard deviation of 16\% with no evident spatial trend across the sky selection window.

\section{Determining stream membership} \label{sec:membership}

In order to search for new RRL stars in the Pal 5 stellar stream, we construct a probabilistic model of the stream and background stellar distribution (simultaneously) using proper motion $(\mu_1, \mu_2) = (\mu_{\phi_1}\cos\phi_2, \mu_{\phi_2})$ and distance $d$ data for all RRLs in the sky window $215^\circ < \alpha < 255^\circ$, $-15^\circ < \delta < 10^\circ$, where $(\alpha, \delta)$ are right ascension and declination and $\phi_1, \phi_2$ are a rotated spherical coordinate system aligned with the stream \citep{Bonaca:2019}.
We \emph{do not} include the known spatial extent or morphology of the Pal 5 stream in our membership model for the stream: We instead later use the sky positions to validate the modeling procedure and selection criteria we then use to define a sample of probable Pal 5 stream RRLs.
This model ultimately enables us to compute stream membership probabilities for RRL stars in a way that fully incorporates \Gaia\ error properties, while simultaneously modeling and marginalizing over uncertainties in the background RRL distribution.

To start, we use the sky track and width of Pal 5 from \citet{Bonaca:2019} to mask out a wide region around the stream---and an extrapolation of the known extent of the stream---in order to define a ``background'' region that should be devoid of Pal 5 member stars (see purple lines in \figurename~\ref{fig:members}).
\changes{We use 1,301 RRL stars in the background region (the full rectangular sky window excluding the stream track) to construct an error-deconvolved density model in the 3D space of proper motion components and distance.}
We use proper motions and uncertainties from \Gaia---using the provided covariance matrix between proper motion components---and distances and uncertainties from the PS1 RRL catalog (\sectionname~\ref{sec:data}).
We optimize the parameters of a Gaussian Mixture Model (GMM) representation of the (deconvolved) background RRL density using \project{extreme deconvolution} \citep[\acronym{XD};][]{Bovy:XD}.
We first fit the background model (with \acronym{XD}) using between $3$--$12$ mixture components and evaluate the Bayesian Information Criterion (BIC) for the optimal parameters in each case.
From this, we find that the BIC is minimized for 6 mixture components: We therefore use 6 mixture components and fix the GMM model parameters to the \acronym{XD}-optimized values and use this as our background model.
We evaluate the background likelihood of all stars in the full sky region (defined above) and refer to this as $p_{{\rm bg}, n}$ (for the $n$th star) below.

\begin{figure*}[t]
\begin{center}
\includegraphics[width=\textwidth]{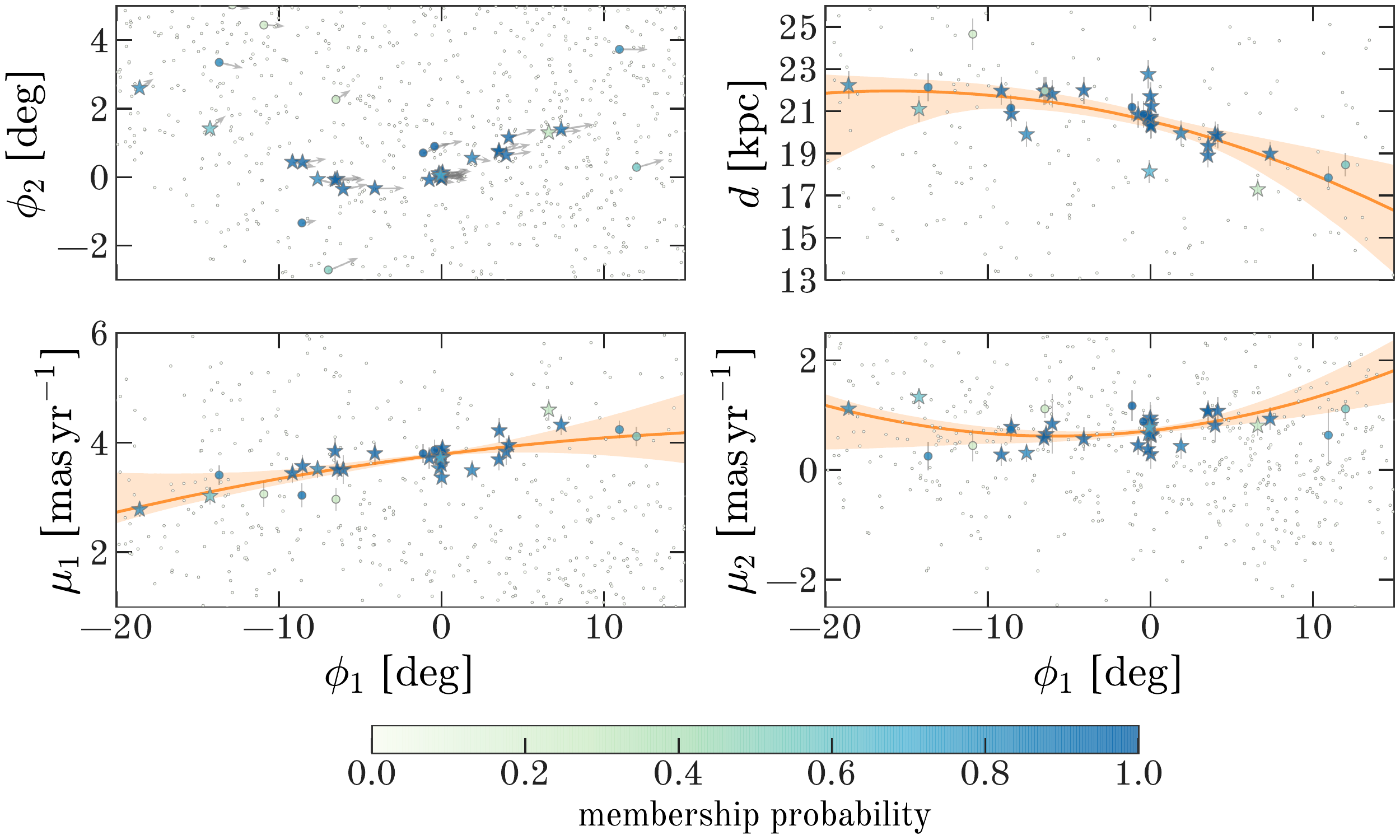}
\caption{RRL stars in the vicinity of the Pal 5 stream colored by membership probability and under-plotted with the inferred stream trends used in the membership model.
We note again that the membership probability does not depend on position or proximity to the stream, and only on distance and proper motions.
\changes{In all panels, RRLs within the stream sky track (see \figurename~\ref{fig:members}) are plotted as star markers, and RRLs with membership probability $<0.1$ are plotted with a smaller marker.}
\textbf{Upper left}: Sky positions (in the Pal 5 stream coordinate frame) of RRLs, with arrows showing the \Gaia\ \DR{2} proper motion vector direction.
\textbf{Upper right}: Distance as a function of Pal 5 longitude, $\phi_1$, for the RRLs. The shaded (orange) band shows the 5--95th percentile confidence region for the inferred distance trend of the stream, and the solid line shows the median posterior sample, all assuming a quadratic function of $\phi_1$.
\textbf{Lower panels}: Same as upper right, but for the proper motion components in the Pal 5 coordinate system, $\mu_1$ and $\mu_2$.
}
\label{fig:trackmembers}
\end{center}
\end{figure*}

To represent the Pal 5 stream, we use a single Gaussian component with a fixed dispersion of $0.05~\masyr$ for the proper motion components and $0.2~\kpc$ in distance.
We have tried including the dispersions as additional parameters in our model but find that they are unconstrained; We therefore fix these values as mentioned to be smaller than the typical uncertainties.
We allow the mean of the Gaussian component to vary independently in each component as a function of $\phi_1$.
Based on the Pal 5 stream models from \citet{Bonaca:2019}, over the range of $\phi_1 \in (-20^\circ, 15^\circ)$, the mean stream trends in $\mu_1, \mu_2$, and distance $d$ can be well-approximated (within our observational uncertainties) by a quadratic function in $\phi_1$.
The mean of our stream component, $\bs{x}$, is therefore given by
\begin{equation}
    \bs{x} = \begin{pmatrix}
        a_{\mu_1} + b_{\mu_1} \, (\phi_1 - x_{\mu_1}) + c_{\mu_1} \, (\phi_1 - x_{\mu_1})^2\\
        a_{\mu_2} + b_{\mu_2} \, (\phi_1 - x_{\mu_2}) + c_{\mu_2} \, (\phi_1 - x_{\mu_2})^2\\
        a_{d} + b_d \, (\phi_1 - x_d) + c_d \, (\phi_1 - x_d)^2
        \label{eq:meanpoly}
    \end{pmatrix}
\end{equation}
where $\bs{\theta} = (a_{\mu_1}, a_{\mu_2}, a_d, b_{\mu_1}, b_{\mu_2}, b_d, c_{\mu_1}, c_{\mu_2}, c_d, x_{\mu_1}, x_{\mu_2}, x_d)$ are free parameters.

Given the above background model, and this model for the stream track, the full likelihood for an RRL star, $n$, with data $D_n = (\mu_1, \mu_2, d)_n$ is represented as a mixture of the two components with mixture weight $f$,
\begin{equation}
    p(D_n \given \bs{\theta}, f) = f \, \mathcal{N}(D_n \given \bs{x}, \mat{C}) + (1-f)\,p_{{\rm bg}, n}
\end{equation}
where $\mathcal{N}(\cdot \given \bs{x}, \mat{C})$ is the multivariate normal distribution with mean $\bs{x}$ and covariance matrix $\mat{C}$.

\begin{figure*}[t]
\begin{center}
\includegraphics[width=\textwidth]{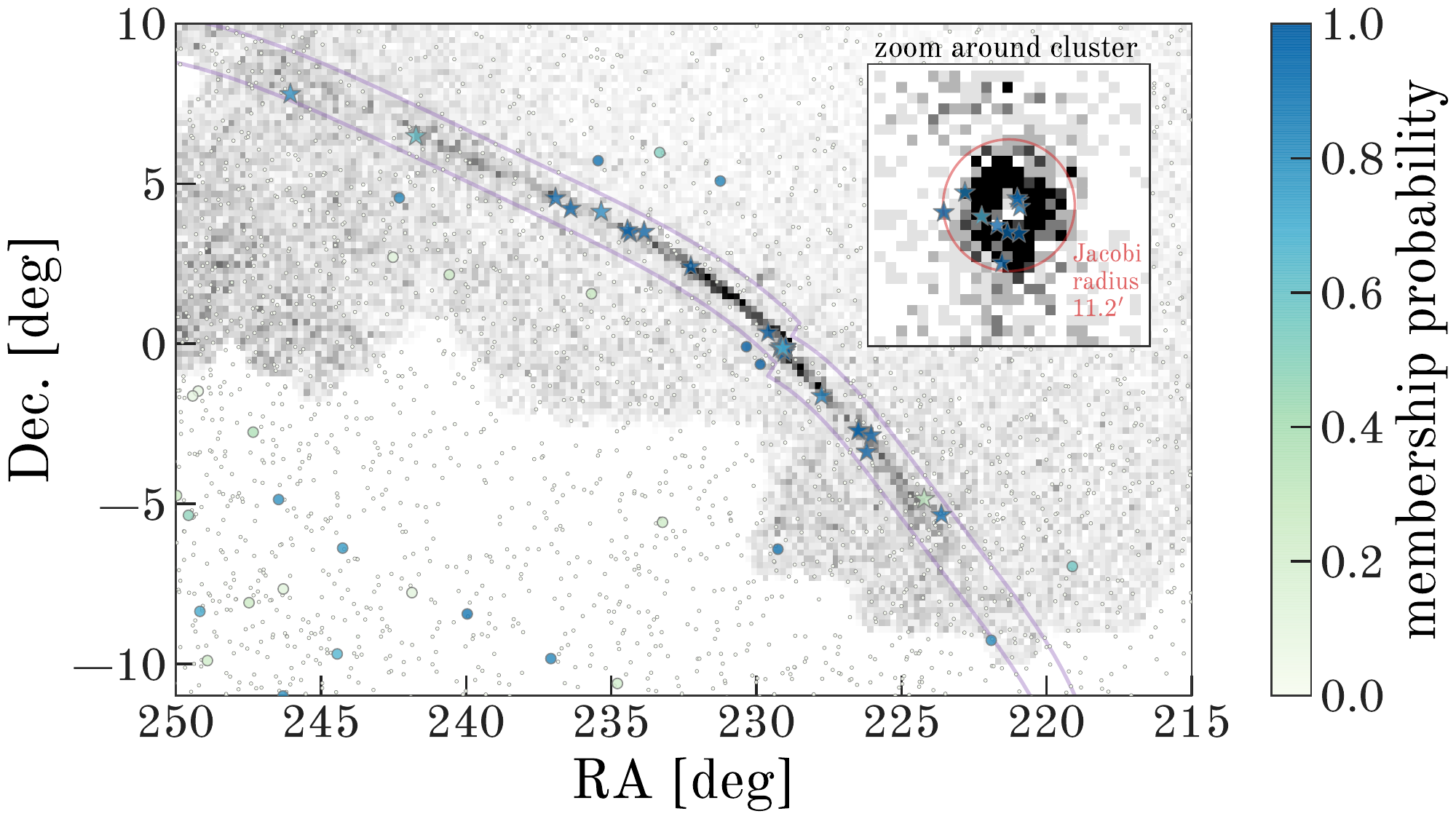}
\caption{Sky positions of RRLs with Pal 5 colored by membership probability.
The background grayscale shows the surface density of CMD-filtered main sequence stars from \citet{Bonaca:2019}.
Note that the membership probability does not incorporate sky position information, but a clear over-density of high-probability RRL members do trace the stream on the sky.
Purple lines outline the stream track window defined in \citet{Bonaca:2019} and used here to exclude stream stars when constructing the background model.
\changes{As with \figurename~\ref{fig:trackmembers}, RRLs within the sky track are shown as star markers, and RRLs with probability $<0.1$ are plotted with a smaller marker.}
The inset panel shows a $1.5^\circ$ by $1.5^\circ$ zoom-in around the Pal 5 cluster, with the Jacobi radius of the cluster shown as a large red circle.
}
\label{fig:members}
\end{center}
\end{figure*}

We use Markov Chain Monte Carlo (MCMC) to generate samples over the parameters $(\bs{\theta}, f)$ using the posterior probability
\begin{equation}
    p(\bs{\theta}, f \given \{D_n\}) \propto p(\bs{\theta}, f) \, \prod_n^N p(D_n \given \bs{\theta}, f)
\end{equation}
where $p(\bs{\theta}, f)$ is the prior probability distribution over the parameters.
We assume that our prior is separable in all of our parameters.
We adopt a uniform prior for the mixture weight, $f$, such that $p(f) = \mathcal{U}(0, 1)$.
\changes{We use Gaussian priors on the mean proper motion and distance using the Pal 5 cluster proper motion measurements from \citet{Vasiliev:2019}, $\bs{\mu} = (-2.728, -2.687) \pm (0.022	0.025)~\masyr$ with a correlation coefficient $\rho_\mu = -0.39$, and the Pal 5 distance from \citet{Kuepper:2015}, $\bs{d} = 23.6 \pm 0.8~\kpc$.
However, we inflate the distance errorbar by a factor of 5 and the proper motion errorbars by a factor of 10 to soften the impact of this prior on our analysis.}

We use broad uniform priors (i.e. $-100 < b, c < 100$ for each coefficient, and $-20^\circ < x < 15^\circ$ for each polynomial centroid) for the trend parameters defined in \equationname~\ref{eq:meanpoly}.
We use the affine-invariant ensemble MCMC sampler \texttt{emcee} \citep{emcee} to sample from the posterior probability distribution given above.
We run \texttt{emcee} with 104 walkers for an initial 1024 steps to ``burn-in'' the sampler, then reset and restart the sampler for another 131072 steps.
We estimate the maximum (over parameters) autocorrelation length using \texttt{emcee}, $\tau_{\rm max} \approx 2200$, and thin the chains by keeping only every 2200th step, leaving us with 6196 posterior samples in our membership model parameters, $\bs{\theta}$.
For each star, we use these samples to compute posterior membership probabilities following \citet{DFM:blog}.

\figurename~\ref{fig:trackmembers} shows a summary of the inferred membership model properties in sky position (shown in Pal 5 coordinates $\phi_1, \phi_2$), distance, $d$, and proper motion components.
\tablename~\ref{t:rrl_members} shows an abbreviated version of our catalog of RRL in the region around Pal 5, with PS1 and \Gaia~ data and membership probabilities appended.
\changes{\tablename~\ref{t:trendpars} contains the maximum a posteriori trend parameter values from the above analysis.}

\section{Results and Discussion} \label{sec:results}

\subsection{RRLs associated with Pal 5 and its tails}
We find a total of \NRRL\ RRLs with membership probability $> 0.5$ that lie within the stream track \citep[again using the track from][see lines in Figure~\ref{fig:members}]{Bonaca:2019}.
Of those, \NRRab\ are classified as type \typeab\ and \NRRc\ are classified as type \typec\ RRLs.
We find a total of \NRRcl\ RRLs within the estimated Jacobi radius of the cluster ($\approx 11'$), meaning that \NRRtails\ RRLs are spread between the leading and trailing tails.
\figurename~\ref{fig:members} shows the sky positions of all RRLs with membership probability $>0.1$ over-plotted on the density of filtered main sequence stars from \citet{Bonaca:2019}:
The RRLs trace the stream track defined by the main sequence stars, but interestingly the density of RRLs does not seem to match the density variations in the under-plotted density map.
In particular, the bulk of the stream RRLs seem to be spread farther out (in angle) from the cluster relative to the main sequence stars, a possible manifestation of the internal dynamics of the initial cluster \citep[e.g., mass segregation][]{Koch:2004}.

We note also that there are a few RRLs outside of the main stream track that appear to be at a consistent distance with consistent proper motions (e.g., the two stars near $\alpha = 230\degr$, $\delta = 0\degr$ in \figurename~\ref{fig:members}) that are not included in the tally above.
No theoretical models of Pal 5 have predicted tidal debris at such locations, so follow-up spectroscopy to confirm or rule out their origin would be very informative.

\subsection{Distance and kinematics of the cluster and stream}

We find a mean Pal 5 cluster distance of \clderr\ using the PS1 RRLs and the membership model described in \sectionname~\ref{sec:membership}.
While past distance measurements to Pal 5 are often reported ranging from $\sim 22.4~\kpc$ to $>23~\kpc$ (e.g., \citealt{Odenkirchen:2009, Erkal:2017}, using measurements from \citealt{Harris:1996, Dotter:2011}), these values are computed from distance moduli that have not been corrected for dust extinction (\citealt{Harris:1996}, and see footnote 5 in \citealt{Fritz:2015}).
Using the recalibrated \citet{Schlegel:1998} dustmap from \citet{Schlafly:2011}, and assuming $R_V=3.1$, we find in the region around Pal 5, $A_V = 0.17~\textrm{mag}$.
Re-deriving distances based on published distance moduli, we find $d_{\rm Pal 5} = 21~\kpc$ \citep{Harris:1996} and $d_{\rm Pal 5} = 20.7~\kpc$ \citep{Dotter:2011}.
Assuming a modest distance uncertainty of $\approx 5\%$ for these isochronal distances, our measurement is consistent with these past measurements.

Using a small number of RRLs then thought to be associated with Pal 5, \citet{Vivas:2006} measured a distance to the cluster of $d = 22.3~\kpc$.
However, this used a now imprecise $M_V$--$\feh$ relation to compute the distance modulus.
Adopting $\feh = -1.35$ \citep{Ishigaki:2016} and using the updated RRL $M_V$--$\feh$ relation from \citet{Muraveva:2018}, we find that this past RRL-based distance to Pal 5 is instead $d_{\rm Pal 5} = 21.2 \pm 0.6~\kpc$.
\changes{Interestingly, \citet{Dai:2018} also independently infer a nearer distance to the cluster around $d_{\rm Pal 5} \approx 21~\kpc$ from dynamical modeling of the stream.}
To summarize: Our inferred distance to the Pal 5 cluster, $d_{\rm Pal 5} = \clderr$, is consistent with (but more precise than) all recent distance measurements, once corrected for dust extinction and updated calibrations of the horizontal branch absolute magnitude.

We find a proper motion $(\mu_{\phi_1}\cos\phi_2, \mu_{\phi_2}) = (3.78, 0.71) \pm (0.02, 0.03)~\masyr$ or $(\mu_\alpha\,\cos\delta, \mu_\delta) = (-2.75, -2.69)~\masyr$, consistent with the measurement in \citet{Vasiliev:2019} within our uncertainties.
The larger (absolute) proper motion (relative to \citealt{Fritz:2015} and other earlier measurements) and the updated distance means that the tangential velocity of Pal 5 is $\approx 16~\kms$ larger than most previous models have assumed.
This is a noteworthy update to consider for future stream modeling efforts and will likely change the inferred Milky Way parameters as compared to past models \citep[e.g.,][]{Kuepper:2015, Bovy:2016}.

We also detect a distance trend from the closer leading tail \changes{($d \approx 19.4~\kpc$ and $\approx$$18~\kpc$ at $\phi_1 \approx 5\degr$ and $10\degr$, respectively) to the more distant trailing tail ($d \approx 21.3~\kpc$ and $\approx$$21.9~\kpc$ at $\phi_1 \approx -5\degr$ and $-10\degr$, respectively)}.
While the proper motion in right ascension appears consistent with constant or only a mild gradient over the extent of the stream, we see a significant gradient in the declination component, with a difference of about $\approx 1~\masyr$ from trailing to leading tail.

\subsection{Properties of the RRL population}

\begin{figure*}[t]
\begin{center}
\includegraphics[width=0.85\textwidth]{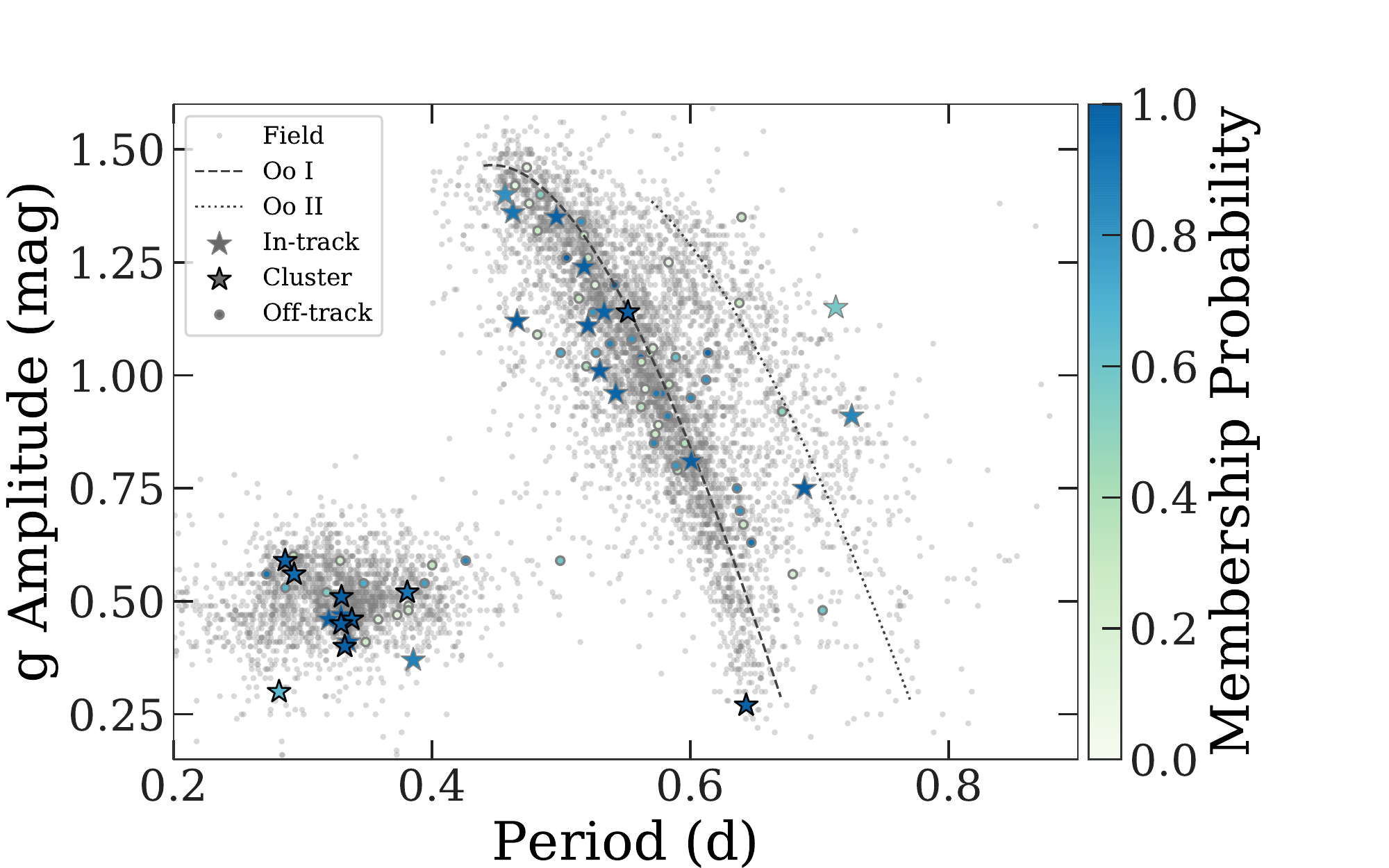}
\caption{Amplitude ($g$-band) versus Period for all RRLs in the Pal~5 field. \changes{Stars with membership probabilities $>0.1$ are colored by membership probability, with those inside the Pal~5 track shown with large (star) markers. RRLs with membership probabilities $<0.1$ are shown as gray markers.} The distribution of the RRLs found in the cluster and its tails is consistent with an Oosterhoff type I (dashed line). }
\label{fig:PA_diagram}
\end{center}
\end{figure*}

Assuming a Jacobi radius of $11'$---computed assuming $\textrm{M}_{\rm Pal 5} \approx 1.2\times10^4~\msun$ \citep{Kuepper:2015} and using the Milky Way mass model in \citet{gala}---we find \NRRcl\ RRLs in the cluster itself (2 type~\typeab\ and 8 type~\typec) and \NRRtails\ in the cluster's tails (13 type~\typeab\ and 4 type~\typec), all with membership probabilities larger than 0.5. In total, we find the number ratio of RRL types to be $N_{ab} / N_{c} \approx 1.25$, a typical ratio found in Oosterhoff I populations \citep{Smith:1995},  comprised mainly of stars still in the zero-age horizontal branch. \figurename~\ref{fig:PA_diagram} shows the period-amplitude diagram for the RRL stars in Pal~5. The full sample of PS1 RRLs is shown in the background for comparison, as a grayscale density histogram. The distribution of the \typeab~RRLs clearly indicate that Pal~5 is an Oosterhoff I cluster, consistent with having a mean period of $0.54~d$. Our findings of the updated ratio of \rrab~to~\rrc, mean period, and Period-Amplitude distribution of the \rrab\ RRLs all support an Oosterhoff I classification for Pal~5, consistent with the cluster's metallicity, $\FeH=-1.4$ \citep{Dotter:2011}.
Out of our newly found \rrab\ RRLs, three are High Amplitude Short Period (HASP, $P <0.48$~d, $A_g>0.75$~mag) stars, again consistent with Pal~5's metallicity, as this type of RRL is only found in relatively metal-rich globular clusters  \citep[$\FeH>-1.5$;][]{Monelli:2017}.

Interestingly, there seems to be a strong segregation of the RRL types: all but two of the \NRRab\ \rrab~are found in the cluster tails and two thirds of the \NRRc\ \rrc~are found in the cluster itself. \changes{
This could be a stochastic effect given the relatively small number of stars, but it seems unlikely given the observed ratios of ab-to-c stars in globular clusters: At the very least, \rrab~stars should account for about half the number of \rrc's \citep[only 5 out of >150 globulars have more than twice as many type~\typec\ than type \typeab\  RRLs, see][]{Clement:2017,Clement:2001,Catelan:2009}. Given the remarkably high completeness of our catalogue, any additional RRL stars would have been detected. Another possibility is that this segregation has a dynamical origin: At fixed metallicity and helium abundance, position along the horizontal branch is determined by (present-day) mass, with less massive stars ending up at higher temperatures as type \typec~stars.
However, if the RRL type segregation had a dynamical origin, the relevant mass should be the one the stars had prior to being ejected from the cluster.} In either case, any of these possible mass differences (either at the main-sequence or the zero-age horizontal branch) are expected to be very small ($\lesssim 0.1~\msun$) and so unlikely to have a significantly different dynamical history---this remains an intriguing puzzle.

\subsection{Stellar Mass and Luminosity}

To estimate the expected total absolute magnitude of Pal 5 we follow the same procedure as in \citet{Mateu:2018}, who use a linear fit of the relation between the absolute magnitude $M_V$ and the number of \typeab\ RRL, based on data from Galactic dwarf galaxies and globular clusters, here including the RR\typec~stars.
The $\log{N_{RR}}-M_V$ relation has a large scatter so, particularly for such low numbers of RRLs, it will only allow us to give rough limits on the cluster's absolute magnitude and, therefore, of its luminosity. Correcting the observed number of 27~RRLs by the median completeness of 92\%, we get a total expected number of 29~RRLs, for which we find a total luminosity $L_V=4.6_{-3.0}^{+8.8}\times 10^4 \mathrm{L}_\sun$.
We then estimate a stellar mass of $0.9_{-0.6}^{+1.8}\times 10^5 \msun$, by assuming a V-band mass-to-light ratio $M/L \approx 2$ \citep{Baumgardt:2019} corresponding to a simple stellar population with the age (12 Gyr) and metallicity ($\FeH=-1.4$) of the cluster \citep{Dotter:2011}, computed from the \citet{Bruzual:2003} models.\footnote{We have used Z=0.0004 ($\FeH=-1.6$), the nearest to the cluster's metallicity available in the models.}
Our mass estimate is larger than that of \citet{Ibata:2017}, who report a current mass of $1.2\pm 0.4 \times10^4\msun$ for the cluster and tails, although their estimate for the initial mass ($4.7\pm 1.5\times10^4\msun$) of the system is at the lower end of our confidence limits.


\subsection{Previously known Pal 5 RRLs}

The Pal~5 cluster has 5 previously known RRL stars (named V1--V5), all of type which are type-\typec, as reported in the \citet{Clement:2001} compilation of variable stars in Galactic globular clusters and dating back to \citet{Kinman:1962}.
In their analysis of the halo density profile with QUEST RRLs, \citet{Vivas:2006} report an overdensity (``Group~6'') that they claim could be associated with Pal~5's tidal tails.
Two of these stars (IDs 403 and 405) are separated by just $\sim10'$ and $11'$ from the cluster's center, which places them within the cluster's Jacobi radius; and an additional star (star 393) outside of the Jacobi radius, but at a similar distance, that could be associated with the stream \citep{Vivas:2006}. We confirm the three stars are highly probable members. \citet{Wu:2005} reports another RRL star (133) as possibly associated, which we also find is a highly probable member. \changes{In summary, nine of the RRL stars in our sample were identified previously as candidate members of the Pal 5 cluster or stream (see the ``Other ID'' column in \tablename~\ref{t:rrl_members}).} More recently, \citet{Ibata:2017} use a prior iteration of the PS1 RRL catalog \citep{Hernitschek:2016}, which contained \emph{plausible} RRLs, to identify 6 and 12 candidate members of the Pal 5 cluster and tails, respectively, based on sky position and distance alone. However, without a published catalog, we cannot compare our sample to theirs.

\section{Conclusions} \label{sec:conclusions}

Using a catalog of RRLs identified using multi-epoch, multi-band photometry from the Pan-STARRS PS1 survey and cross-matched to \Gaia\ \DR{2}, we construct a probabilistic model to determine membership with the Pal 5 cluster and stream.
We find \NRRL\ RRLs with membership probability $>0.5$ (\NRRab\ type~\typeab\ and \NRRc\ type~\typec) over $\approx 18\degr$ of the stream.
From these stars, we infer the distance to the cluster \clderr\ and detect a distance gradient of $\sim 0.2~\kpc~{\rm deg}^{-1}$ between the trailing and leading tail (between $-5^\circ < \phi_1 < 5^\circ$).
We also, for the first time, detect trends in proper motion along the stream, predicted by models of the stream \citep[e.g.,][]{Pearson:2017}.
We release the full RRL catalog in the region around Pal 5 (containing distance information from PS1 and astrometry from \Gaia) with our derived membership probabilities.
We note that past models of the Pal 5 stream have typically used larger cluster distances and slower proper motion values, meaning that any past model streams will be offset from the measurements and analysis presented here \citep[e.g.,][]{Kuepper:2015, Erkal:2017, Pearson:2017}.
These data will be instrumental for improving and updating dynamical models of Pal 5, both for using Pal 5 to constrain the global matter distribution of the Milky Way and in interpreting the observed complex density structure of the stream.

\acknowledgments

This work was performed in part during the Gaia19 workshop and the 2019 Santa Barbara Gaia Sprint (also supported by the Heising-Simons Foundation), both hosted by the Kavli Institute for Theoretical Physics at the University of California, Santa Barbara. The Flatiron Institute is supported by the Simons Foundation. CM acknowledges funding from the MIA program at Universidad de la Republica, Uruguay, and is grateful for the hospitality and support of the Flatiron Institute where part of this research was carried out.

This work has made use of data from the European Space Agency (ESA) mission
{\it Gaia} (\url{https://www.cosmos.esa.int/gaia}), processed by the {\it Gaia}
Data Processing and Analysis Consortium (DPAC,
\url{https://www.cosmos.esa.int/web/gaia/dpac/consortium}). Funding for the DPAC
has been provided by national institutions, in particular the institutions
participating in the {\it Gaia} Multilateral Agreement.

\software{
    astropy \citep{astropy, astropy:2018},
    dustmaps \citep{dustmaps},
    emcee \citep{emcee},
    gala \citep{gala},
    IPython \citep{ipython},
    matplotlib \citep{mpl},
    numpy \citep{numpy},
    scipy \citep{scipy}
}

\clearpage
\appendix

\section{Notes on individual stars}


Star \verb+source_id=6339498589346000768+ is reported both in PS1 and \Gaia~as an \rrc~, with a period 0.4714315854~d in PS1 and 0.320110123108~d in \Gaia~SOS, the two are a pair of 1~d aliases \citep[see e.g.]{Lafler:1965}. In what follows we adopt the shorter period reported in \Gaia~as the correct one as it is the more probable one for an \rrc~star.

Star \verb+source_id=4418731829516302848+ is reported in PS1 and \Gaia~as an \rrab~with a period around 0.646~d and similarly low amplitudes $A_G=0.2$ and $A_g=0.3$ in both surveys. This star is reported in CRTS \citep{Drake:2017} (ID CSS\_J151623.2-000831) as an \rrc~with a period 0.28164~d. We adopt the CRTS period for this star in what follows, as it is based on many more observations ($\gtrsim200$) than \Gaia~and PS1's (about a dozen per filter in each case) and the \rrc~classification and short period are more likely for such a short amplitude star.

\begin{table}
\begin{center}
\caption{Maximum a posteriori proper motion and distance trend parameters (see \sectionname~\ref{sec:membership}). The constant terms have units $\masyr$ and $\kpc$ for proper motion and distance parameters, respectively. The angular units are in degrees such that, e.g., the linear coeff $b_d$ has units of $\kpc~\textrm{deg}^{-1}$.}\label{t:trendpars}
\begin{footnotesize}
\begin{tabular}{c c c c c}
\toprule
y & $x_{\textrm{y}}$ & $a_\textrm{y}$ & $b_\textrm{y}$ & $c_\textrm{y}$\\
param. name & reference $\phi_1$ & constant & linear coeff. & quadratic coeff.\\
\midrule
$\mu_1$ &  $-1.072$ & 3.740  & $4.102 \times 10^{-2}$  & $-6.423 \times 10^{-4}$ \\
$\mu_2$ & $-10.954$ & 0.686  & $-2.826 \times 10^{-2}$ & $2.832 \times 10^{-3}$ \\
$d$     & $-16.081$ & 22.022 & $9.460 \times 10^{-3}$  & $-6.327 \times 10^{-3}$\\
\bottomrule
\end{tabular}
\end{footnotesize}
\end{center}
\end{table}

\begin{table*}[t]
\caption{RRL members of the Pal 5 stream.}\label{t:rrl_members}
\begin{footnotesize}
\begin{threeparttable}[t]
\begin{tabular}{llllllll}
\toprule
\Gaia~DR2               & Other & Period & Amp-$g$ & Type & D & member & separation  \\
\verb+source_id+        & ID\tnote{*}   & (d)    & (mag)   &     &(kpc)& prob & (") \\
\midrule
4418920808077110784 & V5 (K62) & 0.337934 & 0.46 & RRc & 20.7 & 1.00 & 2.0 \\
4418920846732620032 &          & 0.643239 & 0.27 & RRab & 21.7 & 0.99 & 2.0 \\
4418914863842345856 & V1 (K62) & 0.293229 & 0.56 & RRc & 20.4 & 0.95 & 2.0 \\
4418726027016125056 & V3 (K62) & 0.329948 & 0.51 & RRc & 20.7 & 1.00 & 3.9 \\
4418725889577171328 & V4 (K62) & 0.286366 & 0.59 & RRc & 20.3 & 1.00 & 4.5 \\
4418731829516302848 &          & 0.281640 & 0.30 & RRc & 18.1 & 0.65 & 4.9 \\
4418913218870688768 & V2 (K62) & 0.332473 & 0.40 & RRc & 20.3 & 1.00 & 5.1 \\
4418734165978521728 &          & 0.380859 & 0.52 & RRc & 22.8 & 0.91 & 7.7 \\
4418724034151291776 & 403 (V04)& 0.551705 & 1.14 & RRab & 21.2 & 1.00 & 9.8 \\
4418732791593809152 & 405 (V04) & 0.329670 & 0.45 & RRc & 20.7 & 1.00 & 11.0 \\
4419052204012341760 & 133 (W05) & 0.530006 & 1.01 & RRab & 20.8 & 1.00 & 44.5 \\
4418142117622280192 & 393 (V04) & 0.462671 & 1.36 & RRab & 20.0 & 0.91 & 118.8 \\
6339498589346000768 &  & 0.320110 & 0.46 & RRc & 18.9 & 0.99 & 217.8 \\
6339499379619987200 &  & 0.329577 & 0.47 & RRc & 19.4 & 1.00 & 218.6 \\
6339478312804685824 &  & 0.465883 & 1.12 & RRab & 19.9 & 1.00 & 242.9 \\
4421078432143578496 &  & 0.520749 & 1.11 & RRab & 22.0 & 1.00 & 246.1 \\
6339398155830632448 &  & 0.688316 & 0.75 & RRab & 19.8 & 1.00 & 258.9 \\
4427253907921168000 &  & 0.496334 & 1.35 & RRab & 21.8 & 0.99 & 362.6 \\
4427220338456828416 &  & 0.600796 & 0.81 & RRab & 22.0 & 1.00 & 384.4 \\
4427234700827469952 &  & 0.517911 & 1.24 & RRab & 21.9 & 1.00 & 392.4 \\
6337233350579111680 &  & 0.542383 & 0.96 & RRab & 19.0 & 0.98 & 450.7 \\
4427646704153765632 &  & 0.385552 & 0.37 & RRc & 19.9 & 0.87 & 456.6 \\
4424705647988505600 &  & 0.335502 & 0.41 & RRc & 20.9 & 1.00 & 513.0 \\
4426221707021159424 &  & 0.533313 & 1.14 & RRab & 22.0 & 0.99 & 550.1 \\
4450058431917932672 &  & 0.712626 & 1.15 & RRab & 21.1 & 0.57 & 857.8 \\
4439927909737905280 &  & 0.724974 & 0.91 & RRab & 22.2 & 0.85 & 1124.3 \\
4445607604551159040 &  & 0.456636 & 1.40 & RRab & 22.3 & 0.81 & 1394.3 \\
\bottomrule
\end{tabular}
\begin{tablenotes}
\centering
     \item[*] References: K62=\citet{Kinman:1962}, V04=\citet{Vivas:2004}, W05=\citet{Wu:2005}
\end{tablenotes}
\end{threeparttable}%
\end{footnotesize}
\end{table*}

\bibliographystyle{aasjournal}
\bibliography{refs}

\end{document}

%% file: defs.tex
\usepackage{microtype}  
\usepackage{amsmath}
\usepackage{amsfonts}
\usepackage{amssymb}
\usepackage{booktabs}

\usepackage{graphicx}
\usepackage{color}
\usepackage{threeparttable}

\newcommand{\sectionname}{Section}
\renewcommand{\figurename}{Figure}
\newcommand{\equationname}{Equation}
\renewcommand{\tablename}{Table}

\newcommand{\project}[1]{\textsl{#1}}

\newcommand{\acronym}[1]{{\small{#1}}}

\newcommand{\given}{\,|\,}

\newcommand{\mat}[1]{\mathbf{#1}}

\newcommand{\msun}{\ensuremath{\mathrm{M}_\odot}}
\newcommand{\kms}{\ensuremath{\mathrm{km}~\mathrm{s}^{-1}}}

\newcommand{\kpc}{\ensuremath{\mathrm{kpc}}}

\newcommand{\masyr}{\ensuremath{\mathrm{mas}~\mathrm{yr}^{-1}}}

\newcommand{\bs}[1]{\boldsymbol{#1}}

\newcommand{\feh}{\ensuremath{{[{\rm Fe}/{\rm H}]}}}

\newcommand{\FeH} {[\mathrm{Fe}/\mathrm{H}]}

\newcommand{\rrab} {\mbox{RR\emph{ab}}}
\newcommand{\rrc} {\mbox{RR\emph{c}}}

\newcommand{\typeab} {\mbox{\emph{ab}}}
\newcommand{\typec} {\mbox{\emph{c}}}

\newcommand{\Gaia} {{\it Gaia}}
\newcommand{\DR}[1]{\acronym{DR#1}}
\newcommand{\G}{{\mathrm{G}}}
